\documentclass[osajnl,twocolumn,showpacs,superscriptaddress,10pt]{revtex4-1} 

\usepackage{amsmath,amssymb,graphicx}
 \usepackage[latin1]{inputenc}
 \usepackage{srcltx}

\begin{document}


\title{High numerical aperture  holographic microscopy reconstruction with  extended $z$ range}

\author{ N. Verrier$^{a,b}$, D. Donnarumma$^a$, G. Tessier$^{c,d}$ and M. Gross$^a$}

\address{$^a$ Laboratoire Charles Coulomb  - UMR 5221 CNRS-UM2
Universit\'e Montpellier
Place Eug\`ene Bataillon
34095 Montpellier, France \\
$^b$ Laboratoire Hubert Curien - UMR 5516-CNRS-Universit\'e Jean Monnet- 18 Rue du Professeur Beno\^it Lauras 42000 Saint-Etienne, France\\
$^c$ Wavefront Engineering Microscopy Group, Neurophotonics Laboratory, CNRS UMR 8250, University Paris Descartes, Sorbonne Paris Cité, 75006 Paris, France\\
$^d$ ESPCI ParisTech, PSL Research University, CNRS, Institut Langevin, 1, Rue Jussieu, F-75005 Paris, France }

%


\begin{abstract}
An holographic microscopy reconstruction method compatible with  high numerical aperture microscope objective (MO) up to NA=1.4 is proposed.
After off axis and reference field curvature corrections, and  after selection of the +1 grating order holographic image,  a phase mask that transforms the optical elements of the  holographic setup into an afocal device is applied in   the camera plane.  The reconstruction is then made by the angular spectrum method. The field is first propagated in the image half space from the camera to the afocal image of the MO optimal plane  (plane for which MO has been designed) by using a quadratic kernel. The field  is then propagated from the MO optimal plane to the object with the exact kernel.
Calibration of the reconstruction is made by imaging a calibrated object like an USAF resolution target for different positions along $z$. Once the calibration is done, the reconstruction can be made with an object located in any plane $z$. The reconstruction method has been validated experimentally with an USAF target imaged with a  NA=1.4 microscope objective. Near-optimal resolution is obtained over an extended  range ($\pm 50~\mu$m) of $z$ locations.
\end{abstract}


\maketitle

N. Verrier, D. Donnarumma, G. Tessier and M. Gross. Appl. Opt. \textbf{54}, 9540--9747 (2015) 
\url{http://dx.doi.org/10.1364/AO.54.009540}

\section{Introduction}

In Digital Holography  a CCD or CMOS sensor camera records the interference
pattern of the object field wavefront with a known coherent
reference beam. This digital hologram is then used to
numerically reconstruct the  image of the object by propagating
back the measured object field wavefront from the hologram to
the objet \cite{schnars1994direct}.
%
%
%
Many reconstruction methods have been proposed for holographic direct imaging, i.e. without microscope objective (MO)  \cite{schnars1994direct,le2000numerical,yu2005wavelength,zhang2004algorithm,picart2008general}, and for  holographic microscopy (i.e. with MO) \cite{ferraro2003ciw,ferraro_2004_recovering,montfort2006npl,colomb2006npl,colomb2006tac,colomb2006apa}.
%
%
%
In most methods, the holographic reconstruction with MO is made by reconstructing the image of the object enlarged by the microscope objective and not from the object itself.
Then the reconstruction is similar to that made in free space.
%
%
Nevertheless the microscope objective changes the phase of the reconstructed image that must be therefore compensated. Usually this phase compensation is done by adding a lens digital  mask  in the camera plane \cite{montfort2006npl}.

Few methods have been proposed for performing the reconstruction with a large numerical aperture microscope objective.
Coulomb et al.  \cite{colomb2006apa}  proposed to add, in the camera  plane, additional Zernike phase corrections  that  are adjusted to optimize the resolution of the reconstruction. These additional phase corrections are only valid for an object located in the plane where the adjustment was made. In other words when the object is moved along $z$, the phase corrections must be recalculated.
%

%
%


In this paper, we propose a reconstruction method that can be used with a high numerical aperture microscope objective over an extended $z$ range. This method allows to propagate the  hologram (i.e. the optical field) in the object half-space
from  the image of the camera to the MO optimal plane (plane in which MO aberrations are minimal), and then from the MO optimal plane to the object.
The calibration, which is independent of the position of the object, consists in determining the position of the optimal plane and calculating the field in that plane. The proposed reconstruction method and the calibration procedure are validated by a test experiment realized imaging an USAF target by using a high numerical aperture (NA = 1.4) microscope objective.
%
%

%

%
\section{Principles of reconstruction with a large aperture microscope objective}

\begin{figure}[h]
  \begin{center}
   \includegraphics[width=8cm]{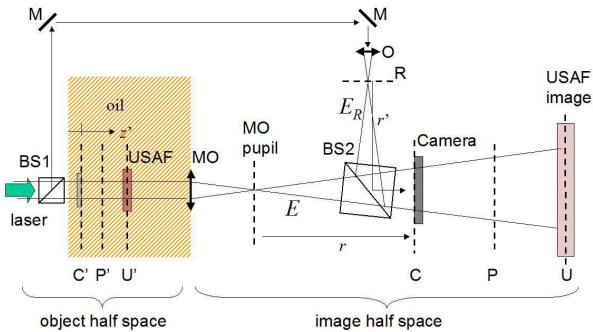}\\
  \caption{Typical holographic microscopy setup. BS1, BS2:
beam splitters; M: mirror. $E$ and $E_R$: signal and reference
optical complex fields; MO: microscope oil objective;   USAF:
USAF target  located in plane U' that is imaged in plane U by  MO; C: camera plane; C': plane of the image of the camera made by MO; P': optimal MO object plane; P: optimal MO image plane; O: short focal lens objective; $r$: MO pupil to camera distance; $r'$: radius of curvature of the reference beam in the camera plane C.  Note that BS2 is angularly tilted in order to
perform off axis holography. Moreover,  the reference beam is a spherical wave whose origin is point R. R is at infinite if the reference is a plane wave. }\label{Fig_figprinciple}
  \end{center}
\end{figure}

Reconstruction with a high numerical aperture oil immersion objective (MO), is illustrated in  Fig. \ref{Fig_figprinciple}. The beam splitter BS2 is angularly tilted to operate in off axis recording geometry.

To get the best possible resolution in conventional imaging  (not holographic),  the object (an USAF target  for instance) and the camera must be  located in the  object and image  optimal planes  P' and P of  the microscope objective MO. This means that the objective  MO has been designed to image  an object located in plane  P' into  plane P. The camera planes (C and C'), the object planes (U and U') and the optimal planes (P and P') must thus coincide (U=P=C and U'=P'=C').


If the imaging is done by digital holography there are less constraints.
To obtain the best resolution with simple reconstruction that involves quadratic phase propagation kernel, it is sufficient that the object is in the optimal plane P'. Thus the condition is U=P and U'=P'. Indeed,
the camera can be located in a plane C that is different from the optimal image plane U=P, as it is still possible to propagate  the field $ E $, from the camera plane C to the  plane U=P of the image of the object.
Since the propagation occurs in free space and since the angles are small, the holographic reconstruction maintains the optimal resolution with quadratic kernel reconstruction.

%
%
%

The proposed reconstruction involves  two steps:

\begin{enumerate}
  \item Field  reconstruction in the optimal plane, with quadratic kernel and correction of the phase,  to obtain the field in  plane P', with the correct  amplitude and phase.
%
  \item Field propagation from the optimal plane P' to the plane of the object U'.
The calculation should be done with the exact propagation kernel, because the propagation takes place in the object half space  with angles that can be large.
%
%
%
%
\end{enumerate}

The reconstruction must be preceded by a calibration procedure to determine:
\begin{itemize}
  \item The location of the optimal plane P',
  \item The phase corrections to be applied to obtain the phase of the  field $E$ in the optimal plane P'.
%
  \item The imaging magnification $G$  to get the pixel size in the optimal plane P', which is needed to calculate the exact propagation kernel.
%
\end{itemize}

\section{Experimental setup}

\begin{figure}[h]
  \begin{center}
   \includegraphics[width=5.5cm]{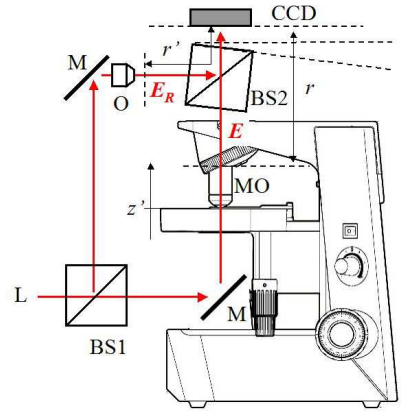}\\
  \caption{Experimental setup. $E_R$, $E$: reference and signal optical fields; L: laser (785 nm); BS1 and BS2: beam splitters; M: mirror; CCD: CCD camera; O: short focal   lens that makes the reference beam divergent; $r$: MO pupil to camera distance; $r'$: radius of curvature of the reference beam in the camera plane C. $z'$: coordinate  of the USAF target plane;  MO: microscope objective that image the USAF target.  }\label{Fig_fig_setup}
  \end{center}
\end{figure}

To illustrate the reconstruction  procedure and to perform calibration, an experimental test has been performed by using the Fig. \ref{Fig_fig_setup} holographic setup.
%

This setup  uses a commercial upright microscope (Olympus CX41) that has been modified; the microscope condenser has been removed and the white light illumination has been replaced by laser illumination. The main laser beam $L$ (Sanyo DL-7140-201: wavelength $\lambda=785$ nm, power 50mW for 95 mA of current)  is split by the beam splitter (BS1) into an illumination beam   and a reference beam.  The object, an U.S. Air Force (USAF) target, is imaged by the microscope objective MO (Nikon: oil, NA=1.4,  $ \times 60$, $160$ mm). The object field $E$ interferes with the reference field $E_R$, and the interference pattern (i.e. the hologram) is recorded by the CCD camera (PCO Pixelfly $1280 \times 1024$ square pixels of size  $\Delta x= 6.7 ~\mu$m). To simplify further digital Fast Fourier Transform (FFT) calculation, the $1280 \times 1024$  matrix measured by the camera is cropped into a $1024 \times 1024$ calculation grid.   In order to perform off-axis holography the beam splitter BS2 that mixes the signal  and reference fields $E$ and  $E_R$ is angularly tilted. Moreover, the reference beam is made divergent by the short focal lens O in order to cover the whole camera detector area. The wavefront of the reference field $E_R$ is thus spherical in the camera plane.
Because the described setup has been built by modifying   a commercial microscope, the optical distance $r$ from camera to MO back focal plane (MO pupil plane)  and the radius of curvature $r'$ of the  reference beam wavefront are not known precisely.

In order to calibrate our setup and to evaluate the performance of the reconstruction procedure, holograms of the USAF  target have been recorded for  $n=1...60$ positions along $z$ i.e. for $z'_n \simeq z'_0 + n \Delta z'$ with  $n=1...60$  and  $\Delta z' = 2.5 \mu$m.
Holograms are recorded with the target located on both sides of camera  plane C' that is  taken as the origin of coordinates $ z' = 0 $.
%
%
This origin corresponds roughly to position $n=29$. Thus we have $z'_{29} \simeq 0$.

%
%
%
%

\section{Hologram reconstruction in the optimal  plane P'.}

\begin{figure}[]
  \begin{center}
   \includegraphics[width=8cm]{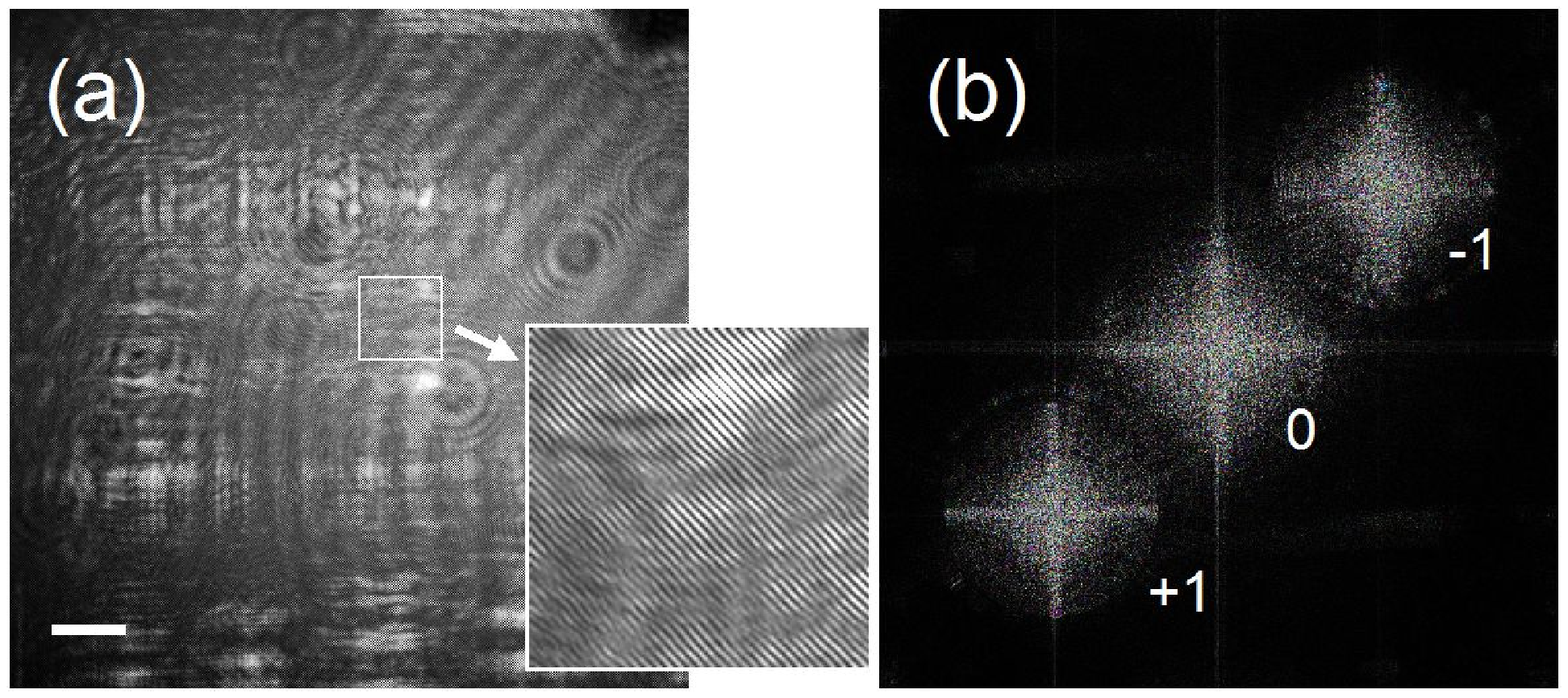}
     \includegraphics[width=8cm]{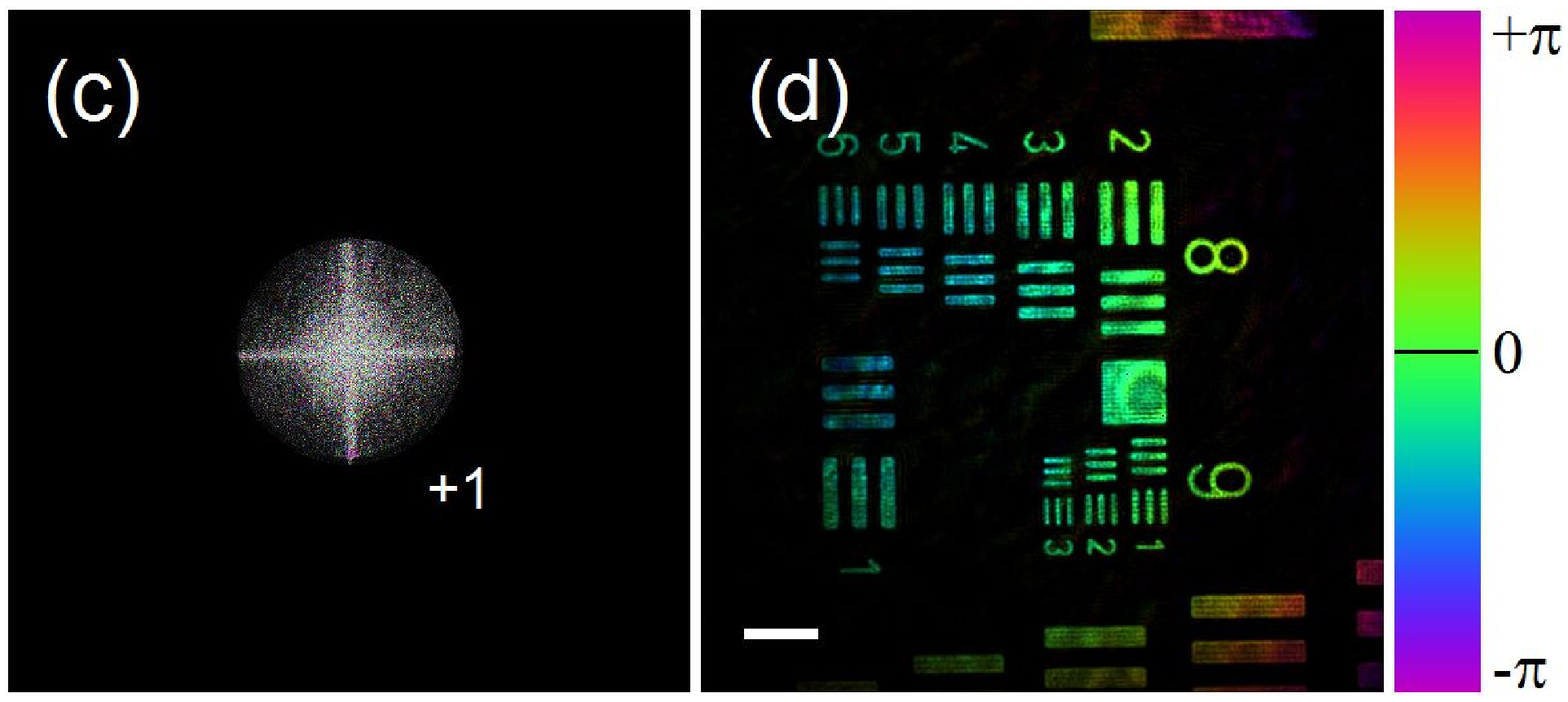}
  \caption{  Holograms ${\tilde H}_{C}$ (a), ${\tilde H}_{2}$ calculated with $dk_x=dk_y=0$ (b),  ${\tilde H}_3 $ calculated with $dk_x/\Delta k= 255$, $dk_y/\Delta k= -244.52$ and $dk_{max}/\Delta k= 162$  (c) and ${ H}_3 $ (d). Holograms are obtained for position $n=1$ (a) and $n=29$ (b,c,d). Arbitrary scale brightness is  ${\tilde H}_{C}$ (a), $|{\tilde H}_{1}|^2$ (b), $|{\tilde H}_{2}|^2$ (c) and  $|{ H}_{2}|^2$ (d); in (d) color is phase i.e.  $\arg( { H}_{2})$. Scale bar is 10 $\mu$m (a,d). } \label{Fig_fig_image_pupil}
  \end{center}
\end{figure}

The hologram recorded in the camera plane C is:
\begin{eqnarray}\label{Eq_H}
  H_C &=& |E + E_R|^2 \\
 \nonumber   &=& |E|^2 + |E_R|^2 + E E_R^* + E^* E_R
\end{eqnarray}
where $E$ and $ E_R$ are the signal and reference fields in the camera plane C. Figure \ref{Fig_fig_image_pupil} (a) shows a typical hologram. Due to the off axis configuration, $ H_C $ exhibits carrier fringes that corresponds to the $E E_R^* + E^* E_R$ terms.
From $H_C$ the hologram $H_{P'}$ in the optimal plane P' has been calculated by using the method we have developed in \cite{verrier2015holographic}. Here, and in the subsequent text, the term ''hologram'' indicates a matrix of complex that is calculated during the reconstruction. In most cases, these ''holograms'' describe the field, or the Fourier transform of the field,  in a plane of the object or image half space.

\begin{figure}[]
  \begin{center}
   \includegraphics[width=8cm]{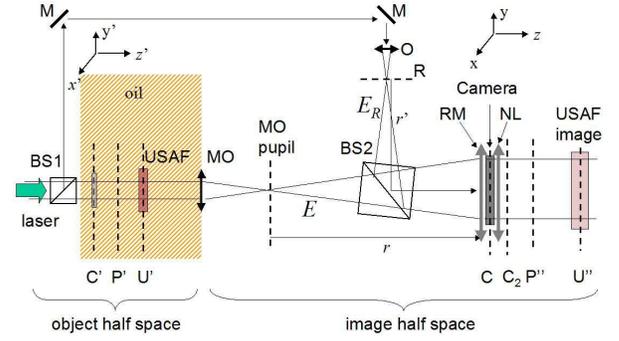}\\
  \caption{
  Reinterpretation of the holographic microscopy setup with the numerical lens NL.
USAF: USAF target located in plane U'; MO  microscope
objective; RM: phase mask located in the camera place C that acts on the reference field $E_R$ to compensate the reference wavefront curvature (focal $r'$) and the off axis tilt; NL: numerical lens of focal $r$ located in the
camera plane C that acts on the signal field $E$ ; MO + NL: afocal optical device that images
the USAF target in plane U'', and the optimal plane P' in plane P''. The  coordinates are $x',y'$ and $z'$ in the object half spaces, and $x,y$ and $z$  in the object half spaces with respect to MO+NL. The  $z'$ and $z$ origins are   planes C' and C.
}
\label{Fig_figprinciple1}
  \end{center}
\end{figure}

The first reconstruction step is to multiply $ H_C$ by a complex matrix $C_{RM}$ which describes a phase mask RM applied on the  reference field $E_R$, whose  purpose  is to compensate for the   wavefront curvature and off axis angle of the reference (see Fig. \ref{Fig_figprinciple1}).    We have:
\begin{eqnarray}\label{Eq_CR_RM}
 \nonumber &&C_{RM}(x,y) = e^{+jk(x^2+y^2)/2r'} e^{j (dk_x  x + dk_y   y) }\\
  &&H_1(x,y)= H_C(x,y)  C_{RM}(x,y)
\end{eqnarray}
where  $k=2\pi/\lambda$,  $\lambda$ the wavelength, and  $x, y$ the transverse coordinates that are discrete quantities whose step is the pixel size $\Delta x=6.7 \mu$m.  In Eq. \ref{Eq_CR_RM},  the phase factors $e^{+jk(x^2+y^2)/2r'}$ is  a lens of focal $r'$ that  modifies the  curvature, and  $ e^{j (dk_x  x + dk_y    y) }$  a prism   that   modifies  the direction of propagation. The parameters $r'$, $dk_x $ and  $dk_y   $ are adjusted so that $E_R$ becomes flat field and propagates on axis, i.e. so that the product $C_{RM}(x,y)  E_R^*(x,y)$ does not vary with $x$ and $y$. The +1 grating order terms of   $ H_1$ (i.e. $ E C_{RM}  E_R^*$) is then proportional to the field $E$ in the camera plane C. Note that the wavefront of $E$ is still curved by MO.

The  second  reconstruction step is to multiply   $ H_1$   by a second matrix $C_{NL}$ describing a numerical lens NL, located in the camera plane, that acts  on the object field $E$ to  compensate for the MO induced wavefront curvature.
\begin{eqnarray}\label{Eq_H2}
 \nonumber &&C_{NL}(x,y) = e^{-jk(x^2+y^2)/2r}\\
  &&H_2(x,y)= H_1(x,y)  C_{NL}(x,y)
\end{eqnarray}
where  $r$ is the focal of NL.
Note that the sign $\pm$ of $j$ in the kernels $e^{+jk(x^2+y^2)/2r'}$  and $e^{-jk(x^2+y^2)/2r}$ of Eq.\ref{Eq_CR_RM} and \ref{Eq_H2} are opposite. Indeed, the first kernel is supposed to act on $E^*_R$, while the second kernel to act on $E$.
The focal $r$ is adjusted so that MO forms with NL an afocal optical device. Note that for a typical microscope
objective MO, the image half space focus plane
coincides with the MO pupil, and $r$ is equal to the distance between the pupil and the
camera, as shown on Fig. \ref{Fig_figprinciple1}.

%
%
%
%

Introducing a numerical lens NL that transforms the
optical elements of the holographic setup into an afocal device
is one of the key points of the proposed reconstruction
method. This afocal device  simplify
the reconstruction. Indeed, all the planes that are conjugated
by an afocal device have the same phase, and are
imaged with the same transverse gain $G$. This gain $G$  is equal to the  imaging gain from plane C' to C (with or without NL). Moreover, the longitudinal
gain $G_L$  remains the same and is finite. Thus, all the conjugated planes of the image half space remains at finite distance $z$. In our test in which an oil immersion objective is used, we have that $G_L=G^2/n_m$, where $n_m$ is the oil refractive index.

The +1 grating order term of   $ H_2$ (i.e. $C_{RM} E \times C_{RM}  E_R^*$) is  proportional to the field $E$ in a plane C$_2$ located just after the afocal device.  This +1 term is selected by spatial filtering in the Fourier space. We must thus calculate  the Fourier space hologram ${\tilde H}_2(k_x,k_y)$:
\begin{eqnarray}\label{Eq_H_2tilde}
  {\tilde H}_2(k_x,k_y) &=& \textrm{FFT} \left[ H_2(x,y) \right]\\
\nonumber    &=& \textrm{FFT}\left [ H_C(x,y) e^{-jk(x^2+y^2)/2r''} e^{j (dk_x  x + dk_y    y) } \right]
\end{eqnarray}
where FFT is the 2D discrete Fourier transform, and  $1/r''=1/r - 1/r'$.
In Eq. \ref{Eq_H_2tilde}, $x, y, k_x, k_y$ are discrete quantities whose step are  $\Delta x$ for $x$ and $y$, and  $\Delta k= 2\pi/(N \Delta x)$ for $k_x$ and $k_y$, where $\Delta x=6.7 \mu$m is the pixel size, and $N=1024$ the size of the calculation grid.

The hologram $| {\tilde H}_2|^2$ calculated without off axis correction
(i.e. with $dk_x   = dk_y    = 0$) is displayed on Fig. \ref{Fig_fig_image_pupil} (b).
The three grating order (+1, 0 and -1) correspond to the
three bright zones of Fig. 3 (b). Should be highlighted here that the edge of +1 grating order is sharp, as noticed
in \cite{verrier2015holographic}. Indeed, since the NL focal length is equal to the
camera-pupil distance $r$, ${\tilde H}_2$ is the reconstructed image
of the MO pupil, made by the Schnars et al. method \cite{schnars1994direct}
 that involves one FFT.

It is good to notice that ${\tilde H}_2$ depends on $r''$ and not on $r$ and $r'$. In section \ref{section_calibration} will be shown that $r''$ can be determined by the calibration procedure, while $r$ and $r'$ cannot.
Thus, using a plane wave reference ($r'=\infty$) does not simplify our reconstruction, since
a spherical reference ($r'\ne \infty$) yields similar calculations. Note also that a spherical reference with
$r'\simeq r$ can be advantageous. Indeed, if
$r = r'$, the MO pupil is located in the lensless Fourier digital
holography plane, and $r''=\infty$. The  reconstruction
of the +1 and -1 images of the pupil is then made by  FFT without kernel. The +1 and -1 images are then both sharp, and so them can be separated more easily.

The prism parameters $dk_x $ and $  dk_y    $
are  adjusted in the way that the pupil image is translated  into
the center of the Fourier space calculation grid, the Fourier space  translation being equal to $dk_x/\Delta k , dk_y/\Delta k$ in pixels Units.
A  circular
crop of the  +1  image of the MO pupil is then made. Since the image of the pupil is sharp, the
spatial filtering made by the crop is optimal \cite{verrier2015holographic}.  The holograms
${\tilde H}_3$ and $H_3$ obtained with this procedure are:
\begin{eqnarray}\label{Eq_tilde_H_3}
 \nonumber {\tilde H}_3(k_x,k_y) &=&  {\tilde H}_2(k_x,k_y ) ~~\textrm{if}~~  \sqrt{k_x^2+k_y^2}<k_{max} \\
\nonumber  &=&0 ~~~~~~~~~~~~~~ \textrm{if not} \\
H_3(x,y)&=&\textrm{ FFT}^{-1} {\tilde H}_3(k_x,k_y)
\end{eqnarray}
In Eq. \ref{Eq_tilde_H_3},   the radius of the selected zone is $k_{max}/\Delta k $ in pixel units. The hologram $| {\tilde H}_3|^2$ calculated by Eq. \ref{Eq_tilde_H_3} with proper  off axis correction ($dk_x/\Delta k= 255$ and $dk_y/\Delta k= -244.52$) and circular crop ($k_{max}/\Delta k =162$) is displayed on Fig. \ref{Fig_fig_image_pupil} (c).

$H_3$ is equal to +1 grating order of  $H_2$ and so is proportional to the field $E$ in plane C$_2$, whose phase is the same than in plane C' (that is conjugated with C$_2$ by the afocal device). This point is illustrated in Fig. \ref{Fig_fig_image_pupil} (d) that shows  $\arg( { H}_{3})$ with colors. Since the control experiment is made with plane wave illumination oriented in the optical axis direction, the phase in planes C' and C$_2$ is flat.

The hologram  $ H_{P''} $ in plane P'', which   is the image of the optimal plane P'  made by MO+NL, is then calculated from $H_3$ by propagating the field in the image half space from plane C (or C$_2$)  to plane P''. Since P'' is at finite distance, this propagation is  calculated by the angular spectrum method that involves 2 FFTs \cite{le2000numerical,yu2005wavelength}. This can be resumed by the following equation:
\begin{eqnarray}\label{Eq_H_P''}
\nonumber H_{P''}(x,y,z_{P''}) &=& \textrm{FFT}^{-1} \left[  e^{j (k_x^2 + k_y^2)z_{P''}/2k} \textrm{FFT}(  H_3(x,y)) \right]\\
 &=& \textrm{FFT}^{-1} \left[  e^{j (k_x^2 + k_y^2)z_{P''}/2k} {\tilde H}_3(k_x,k_y) \right]
\end{eqnarray}
%
Since the origin of the coordinates $z'$ and $z$ are   in planes C' and C,
the reconstruction distance $z_{P''}$ from C to P'' is the coordinate of plane P''. This distance is also equal to $z_{P''}=z'_{P'}G_L$, where $z'_{P'}$ is the coordinate of the object optimal plane P', because of the afocal device. The quadratic kernel   $e^{j (k_x^2 + k_y^2)z_{P''}/2k}$ describes here the propagation in air from C to P''. In Eq. \ref{Eq_H_P''} like in previous equations
(i.e Eq
\ref{Eq_CR_RM}, \ref{Eq_H2}, \ref{Eq_H_2tilde}
and \ref{Eq_tilde_H_3} )
$x, y, k_x, k_y$ are discrete quantities whose steps are  $\Delta x$ for $x$ and $y$, and  $\Delta k$ for $k_x$ and $k_y$. The quadratic kernel of Eq. \ref{Eq_H_P''} can be replaced by the exact kernel $e^{j k_z z }$ with $k_z= \sqrt{k^2 - k_x^2 -k_y~^2}$, but since  $k_x, k_y < N \Delta k \ll k$, the two kernel are equivalent.

Note that here the optimal plane P'' (and not P) must be taked in account,
because this plane is conjugate with P' by the  afocal device
and so it has the same phase. Note also that the
optimal planes P of Fig. \ref{Fig_figprinciple} and P'' of Fig. \ref{Fig_figprinciple1} are very different.
To illustrate this point, imagine to have a microscope
objective designed for infinite distance correction. Plane
P is thus located at $\infty$. The numerical lens NL located
in plane C plays then the role of a tube lens of focal $r$ that
moves the optimal plane from P (at  $ \infty$) to P'', whose coordinate is   $z_{P''}=|\textrm{C P''}|=r$.


%
Because of the afocal device, the  hologram (or the field)  $H_{P''}$ is the exact image of the hologram $H_{P'}$ in optimal plane P' (having the same amplitude and the same phase) but, due to the   MO+NL gain $G$, it will be:
\begin{eqnarray}\label{Eq_H_P''_P'}
&&H_{P'}(x',y',z'_{P'}) = H_{P''}(x,y,z_{P''})\\
\nonumber &&\tilde {H}_{P'}(k'_x,k'_y, z'_{P'}) = H_{P''}(k_x,k_y,z_{P''})
\end{eqnarray}
with $x'=x/G$, $y'=y/G$, $k'_x= G k_x$  and  $k'_y= G k_y$.
Similarly, the hologram $H_3$  in plane C (or C$_2$) is  the exact image of the hologram $H_{C'}$ in the camera image plane C'. It will results also that:
\begin{eqnarray}\label{Eq_H_2 H C'}
&&H_{C'}(x',y') = H_{3}(x,y)\\
\nonumber &&{\tilde H}_{C'}(k'_x,k'_y)={ \tilde H}_{3}(k_x,k_y)
\end{eqnarray}
Thus equation  \ref{Eq_H_P''} can be formally rewritten:
\begin{eqnarray}\label{Eq_H_P'}
H_{P'}(x',y',z'_{P'}) &=& \textrm{FFT}^{-1} \\
\nonumber & & ~~~\left[  e^{j ( {k'_x}^2 + {k'_y}^2)z'_{P'}/2k_m} {\tilde H}_C'(k'_x,k'_y) \right]
\end{eqnarray}
where $k_m=n_m k$ is the wave vector in oil. In Eq.\ref{Eq_H_P'},  $x'$, $y'$, $k'_x$ and $k'_y$ are discrete quantities whose steps are  $\Delta x/G$  and  $G \Delta k$. Although  Eq.\ref{Eq_H_P'} seems to describe the propagation of the field from C' to P' in oil; actually it describes the field propagation from C to P'' in air. Thus, the quadratic kernel $e^{j ( {k'_x}^2 + {k'_y}^2)z'_{P'}/2k_m}$ cannot be replaced  by the exact kernel in Eq. \ref{Eq_H_P'}.

\section{Reconstruction of an  object located outside the optimal plane: U' $\ne$ P'.}

 The hologram $ H_{U~'} $ in the USAF plane U'  can be obtained  from $H_{P~'}$  by propagating the field in the object half space from plane P' to plane U'. Here again, the propagation can be calculated by the angular spectrum method \cite{le2000numerical,yu2005wavelength} that involves 2 FFTs.  Nevertheless, since the numerical aperture NA can be large (NA=1.4 in the test experiment), we must use the exact propagation kernel $e^{-j k'_z z'}$, and not the quadratic one $e^{j ({k'_x}^2 + {k'_y}^2)z'/2k_m}$. We get thus:
\begin{eqnarray} \label{Eq_H_U__bis}
  H_{U'}(x',y',z') &=&   \textrm{FFT}^{-1}  \\
\nonumber  &&\left[ e^{j k'_z (z'-z'_{P'})}\textrm{ FFT} \left[ H_{P'}(x',y',z'_{P'}~ \right] ~\right]
\end{eqnarray}
where $(z'-z'_{P'})$ is the propagation distance from plane P' to plane U', and $k'_z$ the $z$ component of the wave vector $\textbf{k}$ in  oil of optical index $n_m$. $k'_z$   is a function of $k'_x$ and $k'_y$. We have:
\begin{eqnarray} \label{Eq_k'_z}
 k'_z&=&\sqrt{k_m^2-{k'_x}^2-{k'_y}^2}
\end{eqnarray}
with $k_m=n_m 2\pi/\lambda$. In equations \ref{Eq_H_U__bis} and \ref{Eq_k'_z}, $x', y'$ and  $k'_x, k'_y$ are discrete quantities whose steps are $\Delta x/G$ and  $G \Delta k$.
We can combine Eqs. (\ref{Eq_H_P'}) and (\ref{Eq_H_U__bis}) to get:
\begin{eqnarray} \label{Eq_H_U ter}
  &&H_{U'}(x',y',z') = \textrm{FFT}^{-1}      \\
\nonumber &&~~~~ \left[  e^{j k'_z (z'-z'_{P'})}   e^{j ({k'_x}^2 + {k'_y}^2)z'_{P'}/2k_m}  \textrm{ FFT} \left[ H_{C'}(x',y')~ \right] \right]
\end{eqnarray}
This equation  summarize the reconstruction  from $H_{C'}$ to $H_{U'}$ in a very compact and useful way. This equation is nevertheless difficult to interpret since the two kernels describes two different physical process. The first  kernel  $e^{j ({k'_x}^2 + {k'_y}^2)z'_{P'}/2k_m} $ is quadratic. It describes the propagation of the hologram in air from C to P''. Indeed, for the discrete coordinates $k'_x, k'_y$ that are used in the  calculation,  $e^{j ({k'_x}^2 + {k'_y}^2)z'_{P'}/2k_m} $ is exactly equal to the kernel  $e^{j ({k_x}^2 + {k_y}^2)z_{P''}/2k} $ that describe the C to P'' propagation.  On the other hand, the second kernel $ e^{j k'_z (z'-z'_{P'})} $ describes the propagation  in oil from P' to U'.


%
Le us summarize the Eq. \ref{Eq_H_U ter} reconstruction:
\begin{itemize}
  \item The field is first propagated from the camera image plane C' to the  optimal plane  P' by using the quadratic kernel $e^{j ({k'_x}^2 + {k'_y}^2)z'_{P'}/2k_m}$,  which describes the propagation of the field in the image half space  from  C (or C$_2$) to P''.
%
%
  \item The field is then propagated from the optimal plane P' to the  object plane  U'. As this propagation is done in the object half-space with a high numerical aperture, we must  use the exact kernel  $ e^{j k'_z (z'-z'_{P'})}$.
%
\end{itemize}

\section{Calibration of the experimental setup.}\label{section_calibration}
%
%

Before performing the reconstruction, it is necessary to calibrate the setup. This calibration consists of the determination of the parameter $ r''$ of  Eq. \ref{Eq_H_2tilde}, the prism parameters  $dk_x , dk_y   $ of   Eq. \ref{Eq_CR_RM} and \ref{Eq_H_2tilde},  the afocal magnification  $ G $   and the position $z_{P''}$ (or $z'_{P'})$  of the optimal plane P'' (or P').

The calibration of $ G $, $ r'' $ and $ dk_x ,  dk_y   $ is made by using the afocal idea as follows \cite{verrier2015holographic}.
%
\begin{itemize}
  \item The imaging magnification $ G $ is measured by positioning the USAF target  in the direct imaging plane (U'=C') and by measuring the magnification from plane U'=C' to plane U=C. In our test experiment, we got $G=74.64$ yielding an  pixel size of  $\Delta x'=89.8$ nm in the object half space. Note that $G$ is not equal to the  nominal gain of the our objective ($\times 60$).
%
  \item  The parameter $ r''$ is obtained by adjusting $  r''$ so that the size  of the reconstructed image does not depend on the position $z'$ of the USAF target. The  magnification is then  equal to $ G $ for all positions $z' $ of the USAF target and in particular  for $z'=0$ (plane C' to plane C or  C$_2$) and $z'=z'_{P'}$ (plane P' to plane P''). We got $1/r''=-0.58 ~\textrm{m}^{-1}$.
%
%
  \item  The prism parameters  $dk_x  , dk_y    $ are obtained by adjusting  $dk_x$ and $dk_y$ so that the $x', y'$  position of the reconstructed image  does not depend on the position  $z' $ of the USAF target. We got $dk_x/\Delta k = 255 $ and $dk_y/\Delta k = -244.52 $.
%
\end{itemize}

\begin{figure}[h]
  \begin{center}
   \includegraphics[width=7cm]{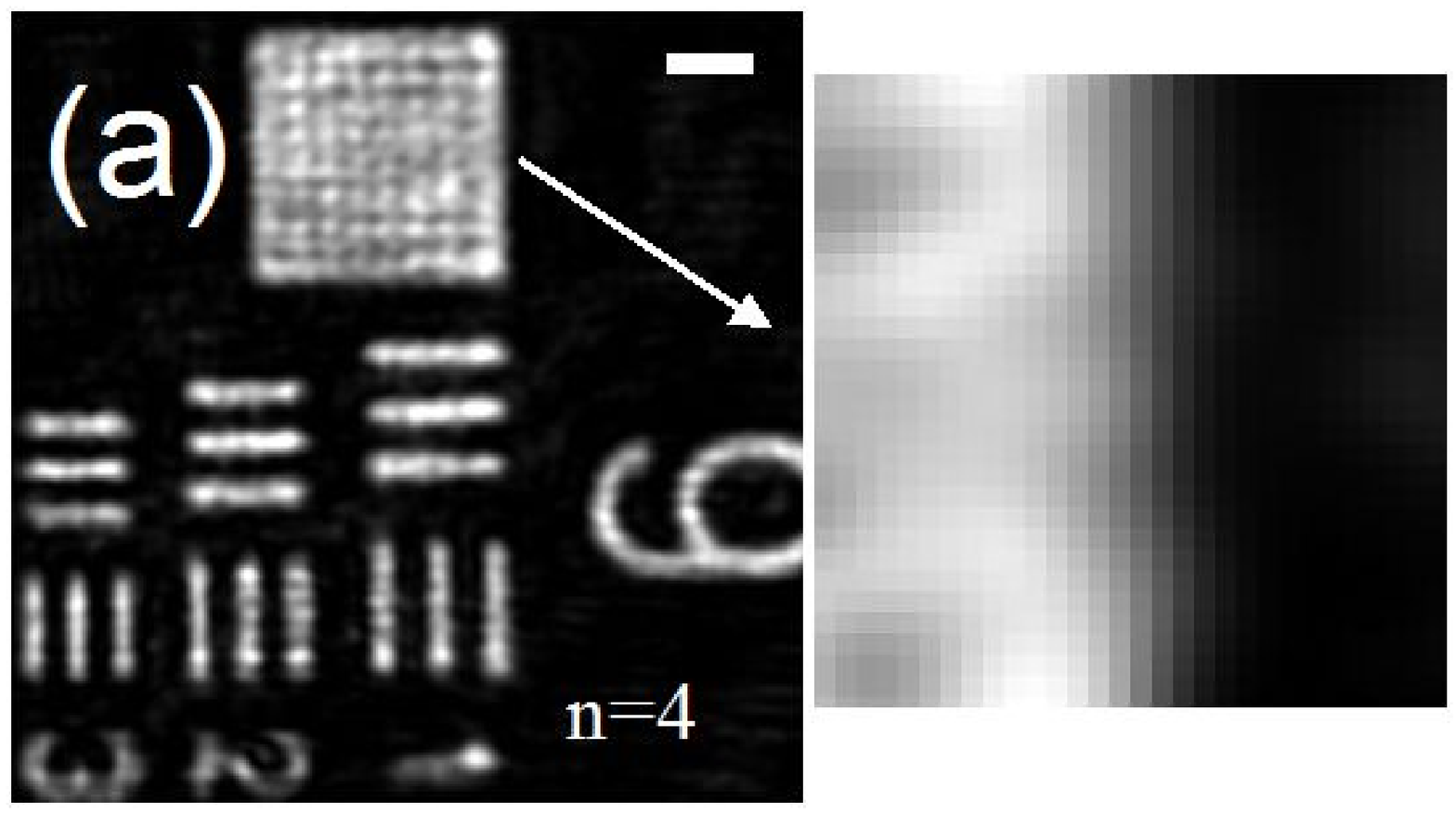}\\
     \includegraphics[width=7cm]{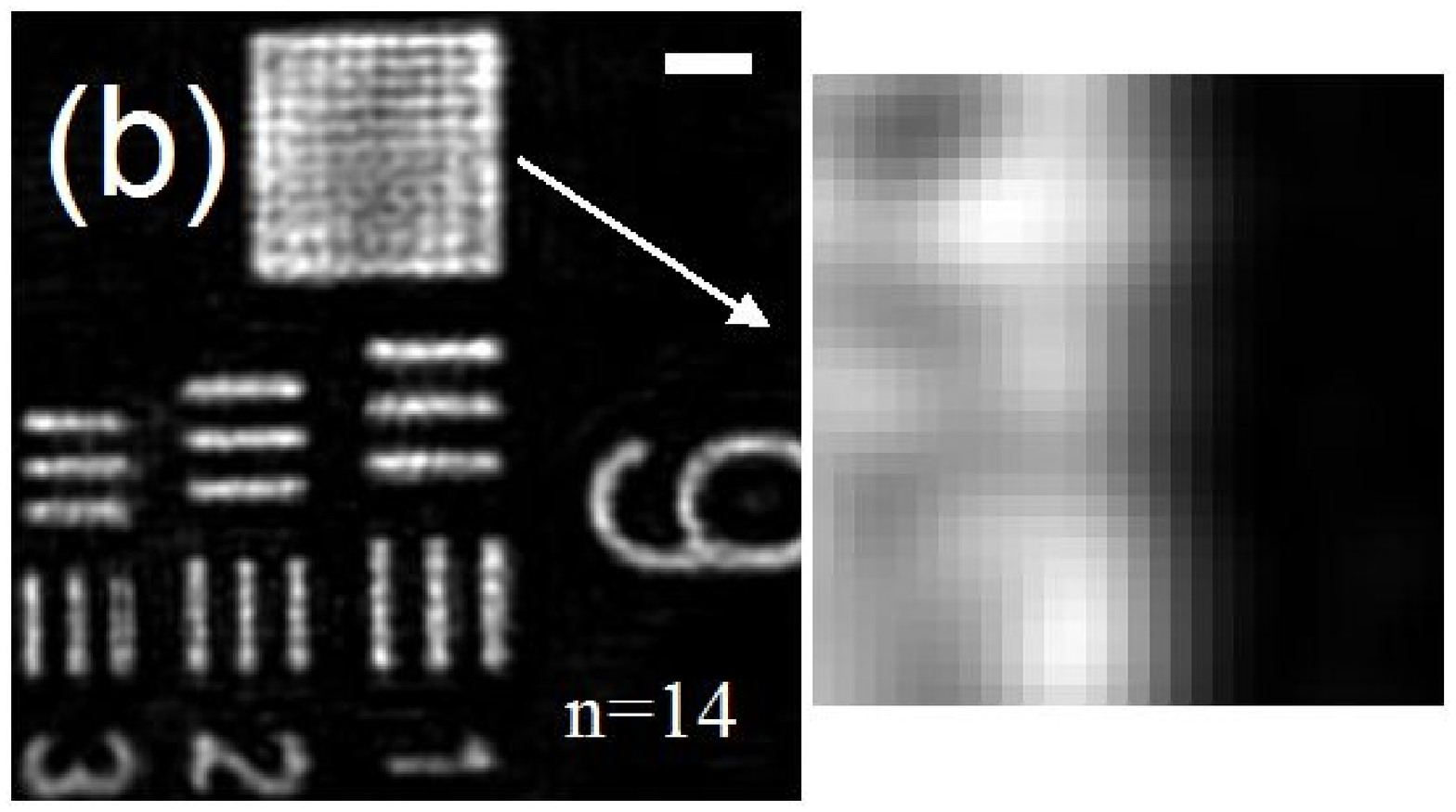}\\
  \includegraphics[width=7cm]{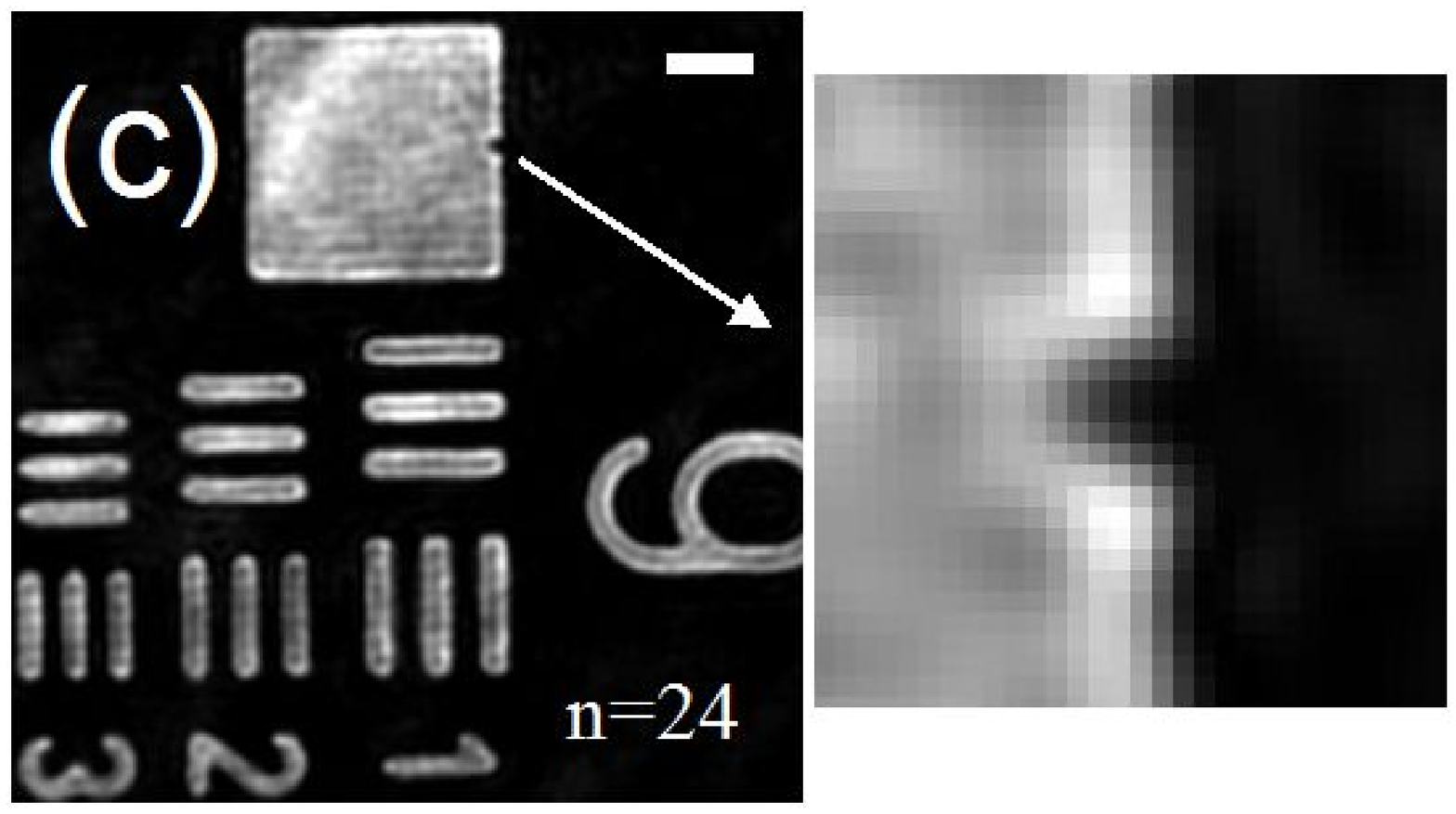}\\
  \includegraphics[width=7cm]{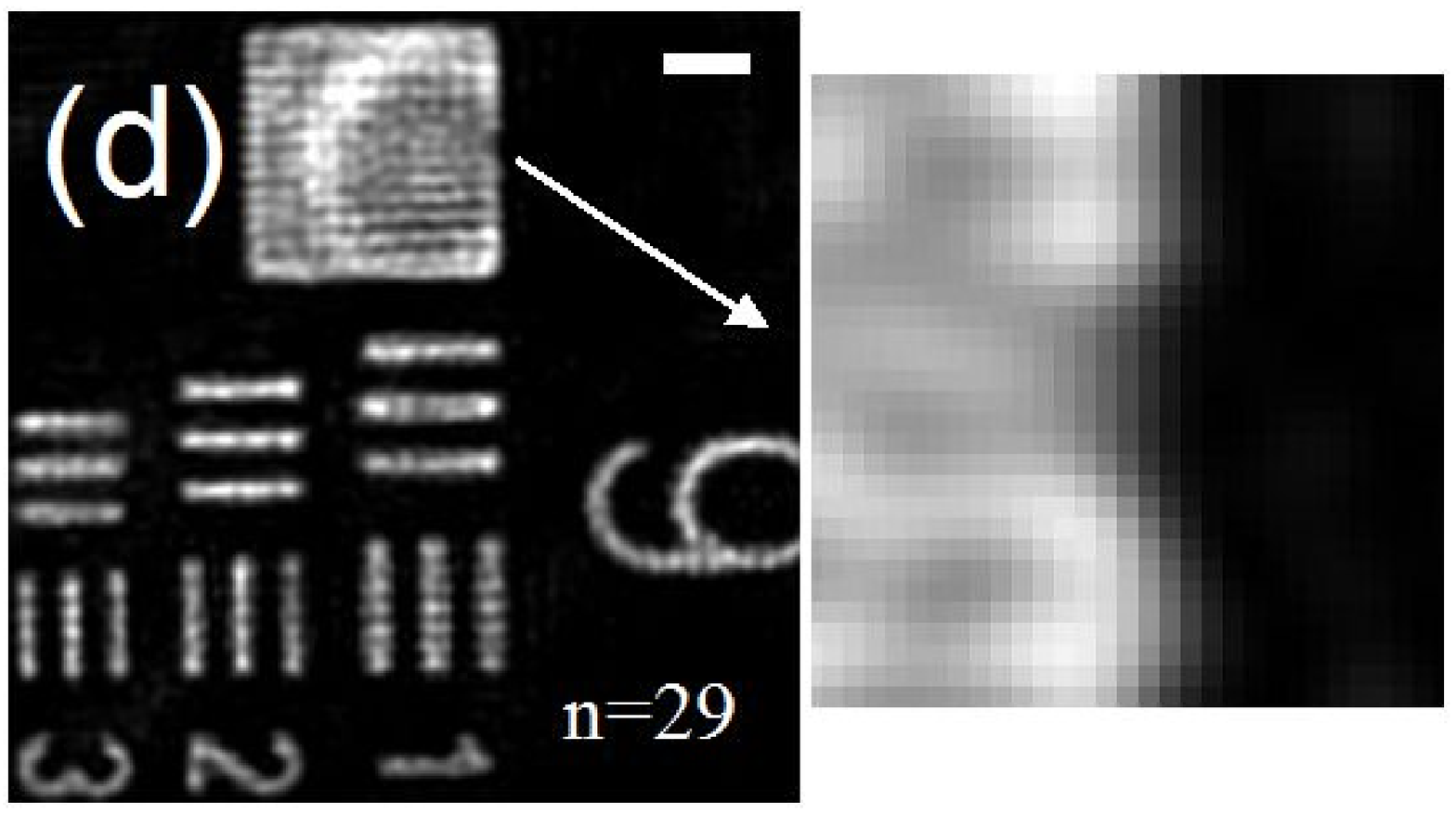}\\
    \caption{ Zooms( $300\times 300$  and $30\times 30$ pixels)  of the center of the reconstructed image $|H_{U'}|^2$  ($1024\times 1024$ pixels) of the USAF target obtained for the locations: $n=4$ (a), $n=14$ (b), $n=24$  (c) and $n=29$  (d). Numerical aperture is NA=1.4.  $H_{U'}$ is calculated  by Eq. \ref{Eq_H_U ter} with $z'_{P'}=0$  by adjusting $z'_n$ to minimize $S$. Calculation is made with $z'_n=-63.69$ (a), -38.54 (b), -13.39 (c) and -0.86 $\mu$m (d). Display is made  in arbitrary linear scale.  Scale bar is 3 $\mu$m, pixel size is $\Delta x'=89.8$ nm.
      } \label{Fig_fig11}
  \end{center}
\end{figure}

The calibration of    the location  ${z'}_{P~'}$ of the optimal plane P' is made by calculating the USAF image with a quadratic kernel and by finding the position that gives the best reconstructed image.
For  all position $n$ of the USAF target, we have  thus calculated  $H_{U'}(x',y',z'_n)$   by  Eq. \ref{Eq_H_U ter}  with $z'_n-z_{P'}=0$ (i.e.with the quadratic kernel only).
For each position $n$, we  have adjusted  $z'_n$ so as to obtain the sharpest image $H_{U'}(x',y',z'_n)$ . As the USAF
target is an ``amplitude object'', we have adjusted $z'_n$ by using the ''focus plane detection criterion'' of
Dubois et al. \cite{dubois2006focus}, which minimizes  $S$:
\begin{eqnarray}\label{Eq_S}
  S &=& \sum_{x',y'} \left| H(x',y', z'_n) \right|
\end{eqnarray}
where $H$ is the hologram that is considered. Here, $H=H_{U'}$.

Figure \ref{Fig_fig12} shows  zooms ($300\times 300 $ and $30\times 30$ pixels) of the reconstructed images ($1024\times 1024$ pixels) obtained after adjustment of $ z'_n$ for  positions $n = 4 $ (a), 14 (b) 24 (c) and 29 (d). The figure illustrates the  effect of the USAF position on the quality of the  holographic  quadratic kernel reconstruction. Position $ n = 24 $ has the best visual resolution and corresponds  to the absolute minimum of the criterion $S$. Indeed, the absolute minimum of $S$ is reached for $n=24$ and $z'_{24}=-13.39 \mu$m.  Position $n=24$ of the USAF target corresponds thus roughly to the optimal plane (i.e. U'=P'), with  $z'_{P'} \simeq z'_{24}=-13.39 ~ \mu$m. When the USAF target is located outside the optimal plane (i.e. for U'$\ne$ P'), the optimal resolution cannot be reached and the resolution is lower  as seen on   Fig. \ref{Fig_fig11} (a,b) and (d) that display the best reconstructed images that have been obtained for  positions $n=4$, 14 and 29.

By selecting the best position $n$,  $z'_{P'}$  is measured with an accuracy that is limited by the USAF $z$ displacement step  (2.5 $\mu$m). To avoid this quantization problem,  we have determined  $z'_{P'}$  by adjusting both  $z'-z'_{P'}$  and $z'_{P'}$   in Eq. \ref{Eq_H_U ter}   so as to minimize $S$. For position $n=24$, we got  $z'_{24} - z'_{P'}=0.8~ \mu$m and  $z'_{P'}=-14.34~ \mu$m.

We have thus consider, here and in the following, that the exact location of the optimal plane P' is $z'_{P'}=-14.34 ~\mu$m.

\section{Reconstruction with a tube lens and a microscope objective corrected at infinite distance. }

\begin{figure}[]
  \begin{center}
   \includegraphics[width=8cm]{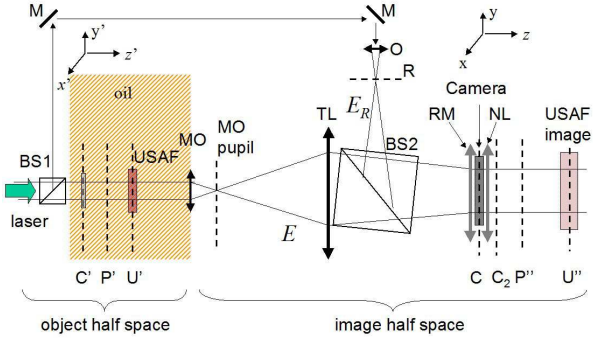}\\
  \caption{
 Holographic microscopy setup with a tube lens TL that is very similar to the Fig.\ref{Fig_figprinciple1} setup.
}
\label{Fig_figprinciple2}
  \end{center}
\end{figure}

All the results presented here have been obtained with an objective corrected a finite distance (150 mm) and without tube lens. There remain nevertheless valid with a tube lens TL and with an objective MO corrected at finite or infinite distance \cite{verrier2015holographic}.
The tube lens is a long focal length  lens located somewhere between the MO pupil and the camera.

Figure \ref{Fig_figprinciple2} shows an example of setup with a tube lens TL located in between the MO pupil and the beam splitter BS2.
The tube lens TL modify the optical arrangement, but it is nevertheless possible to find a numerical lens NL that transforms the ensemble of lenses (MO+TL+ NL) into an afocal device. The  focuses  $r$ and $r'$ of both the numerical lens  NL and the reference mask RM
must be modified to account for the tube length, but this is done by  the calibration procedure, which  adjusts $r''$ so that the USAF target image keeps the same size for all position $z'$. The position $z_{P''}$ of the optimal plane and the afocal gain $G$ are modified too, but the calibration procedure, which is based on the afocal character of the ensemble of lens  (MO+TL+ NL)  remains the same.

\section{Experimental validation of the USAF reconstruction for all $ h $ positions.}

\begin{figure}[h]
  \begin{center}
   \includegraphics[width=7cm]{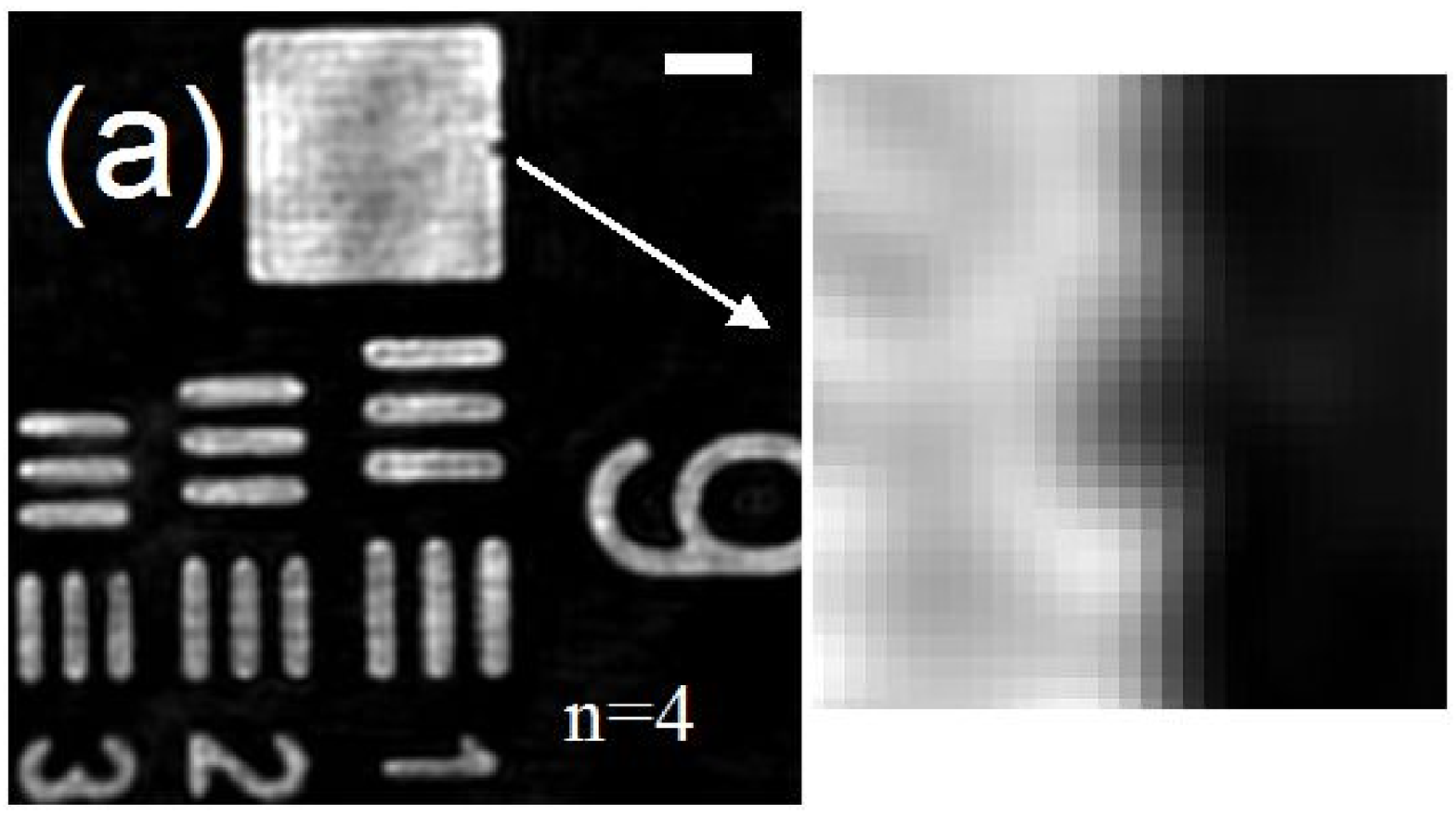}\\
   \includegraphics[width=7cm]{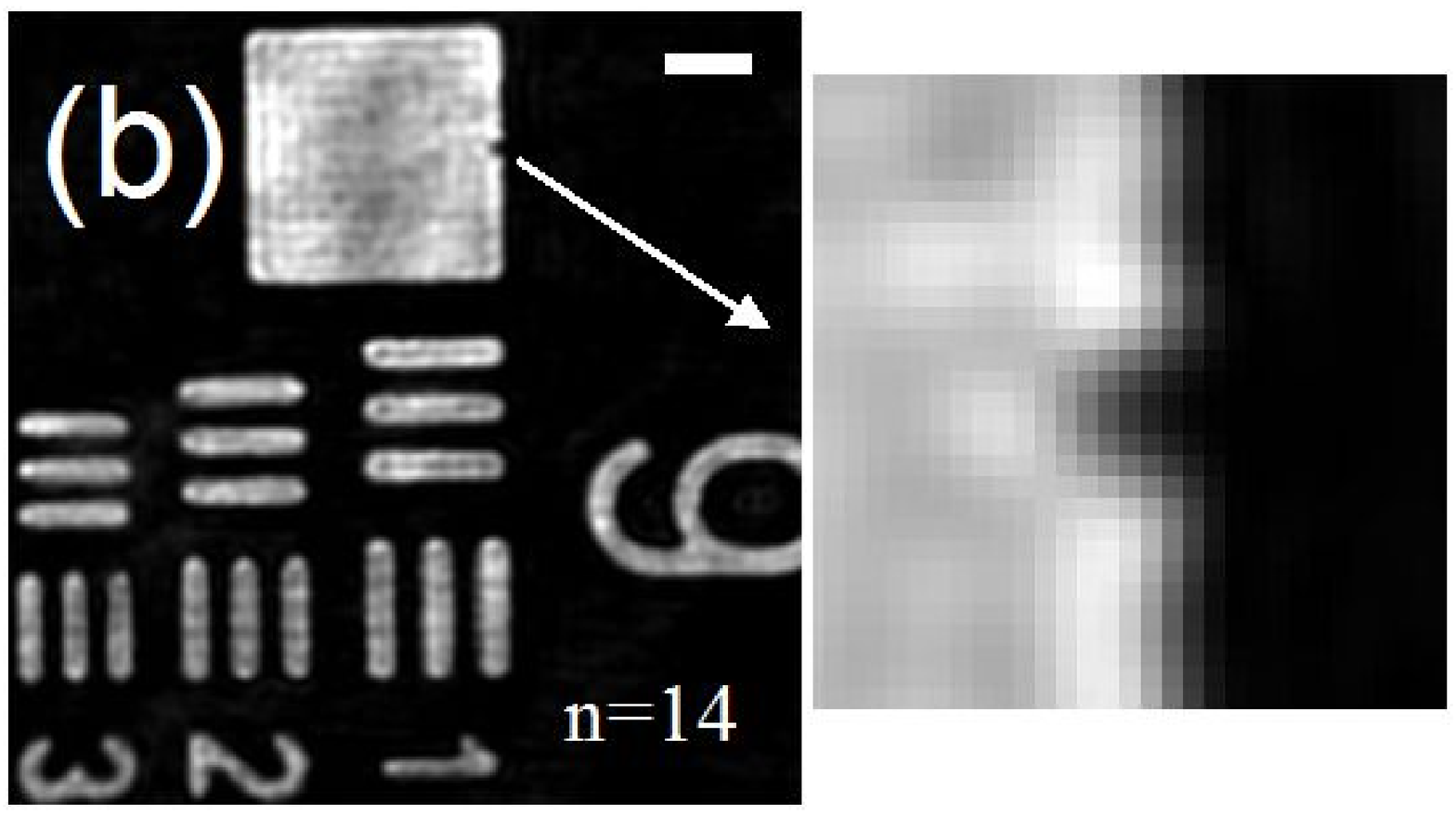}\\
   \includegraphics[width=7cm]{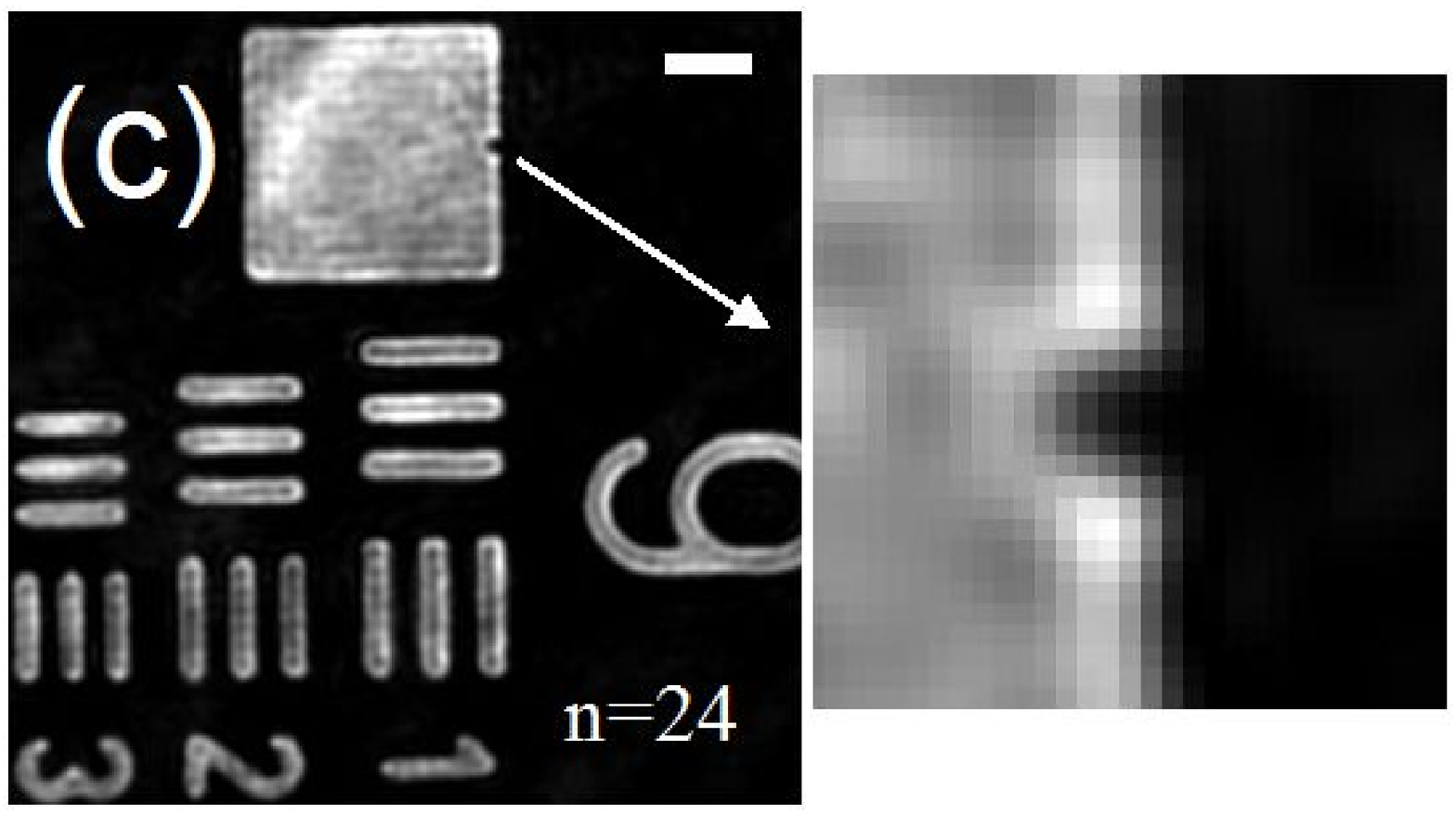}\\
   \includegraphics[width=7cm]{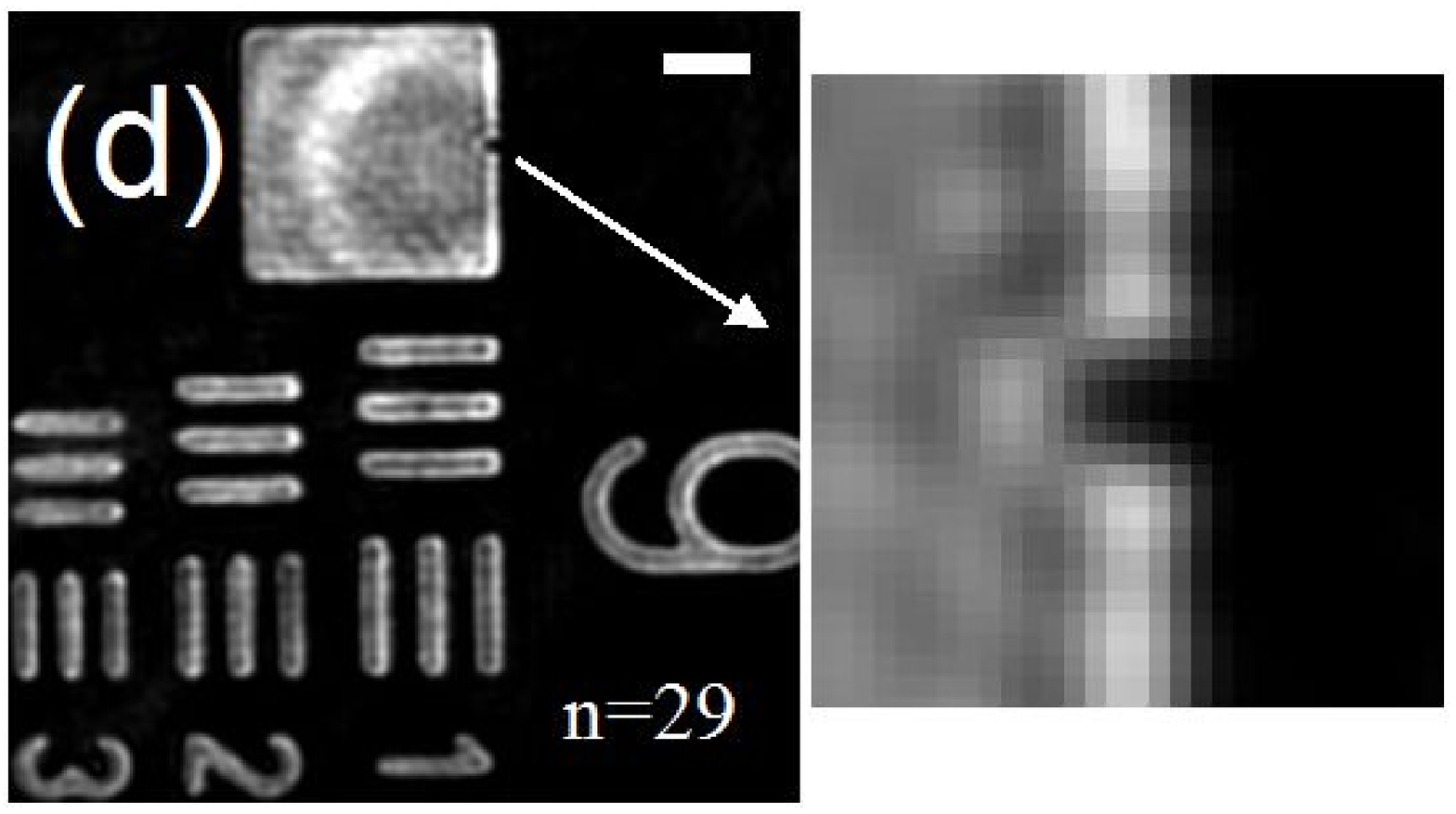}\\
    \caption{ Zoom ($300\times 300$ pixels) of the center of the reconstructed image $|H_{U'}|^2$ ($1024\times 1024$ pixels) of the USAF target obtained for the locations: $n=4$   (a), $n=14$  (b), $n=24$ (c) and $n=29$ i.e. (d). Numerical aperture is NA=1.4.  $H_{U'}$ is calculated with Eq. \ref{Eq_H_U ter} by adjusting $z'_n-z'_{P'}$ and  $z'_{P'}$  to minimize $S$. Calculation is made with $z'_{P'}=-14.34 \mu$m and $z'_{n}-z'_{P'}= -47.69 $ (a), -23.39 (b), 0.8 (c)  and 12.84 $\mu$m (d).   Display is made  in arbitrary linear scale. Scale bar is 3 $\mu$m, pixel size is $\Delta x'=89.8$ nm.} \label{Fig_fig12}
  \end{center}
\end{figure}

To validate the proposed method, we have performed the reconstruction for all positions $ n = 1 ... 60$ of the USAF target  by calculating $ H'_{U'}$ given by Eq. \ref{Eq_H_U ter}. For each $n$ position, we have calculated  $ H'_{U'}$  with $z'_{P'}=-14.34~\mu$m  by adjusting  $z'_n$ in order to  minimize    $S $.

Figure \ref{Fig_fig11} presents the zooms  ($ 300 \times  300$ and $ 30 \times  30$ pixels) of the center of the USAF  images  ($1024\times 1024$ pixels)   reconstructed for  $ n = $ 4, 14, 24 and 29.
To better visualize the resolution,  we have center the $ 30 \times  30$ pixels zoom on a small defect in the bright square zone of the USAF target (white arrow). Note that the  pixels, whose size is $\Delta x'=89.8$ nm,  are visible on the $ 30 \times  30$ zoom.

Figures   \ref{Fig_fig11} and  \ref{Fig_fig12}, allow us to  compare the quadratic kernel simplified calculation made in  \cite{verrier2015holographic} (Fig.  \ref{Fig_fig11}) with the full two kernels calculation of Eq. \ref{Eq_H_U ter} (Fig.  \ref{Fig_fig12}).
%
\begin{itemize}

 \item For $ n = $ 4 and $ n = $ 14, the USAF target  is shifted by $ z'_n-z'_{P'} \simeq $ -50 and  -25 $ \mu$m with respect to the optimal plane P'. The resolutions obtained with the simplified reconstruction are very degraded (Fig.\ref{Fig_fig11} (a) et (b) ), while the ones obtained with two kernels are good (Fig.\ref{Fig_fig12} (a) et (b) ).
  \item For $ n = 24 $, the USAF target is near the optimal plane $z'_{24} \simeq z'_{P'}$. The full and simplified calculations are mainly made with  the quadratic kernel and the images  Fig.\ref{Fig_fig11} (c) and  Fig.\ref{Fig_fig12} (c)  are nearly identical.

  \item For $n = 29 $, the USAF target is shifted by $  z'_{29}-z'_{P'} \simeq  +13 ~\mu$m with respect to  P', and close the camera plane C'.  The simplified reconstruction is thus done nearly without any kernel since $ z'_{29} \simeq $ 0. Even in that case, the simplified reconstruction remains very imperfect (Fig.\ref{Fig_fig11} (d) ), while the full  reconstruction is excellent (Fig.\ref{Fig_fig12} (d)).
\end{itemize}

\begin{figure}
  \begin{center}
   \includegraphics[width=7cm]{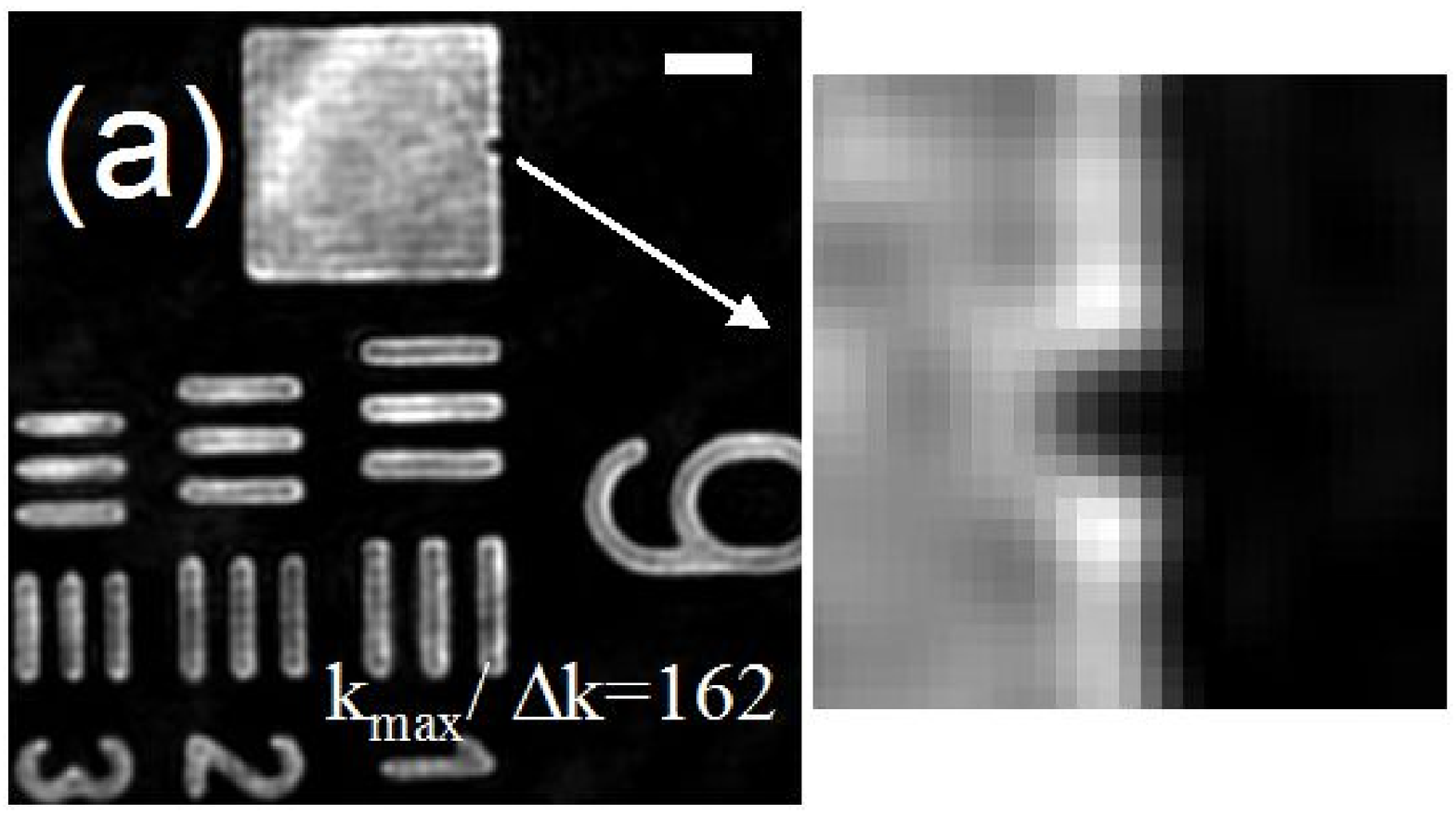}\\
   \includegraphics[width=7cm]{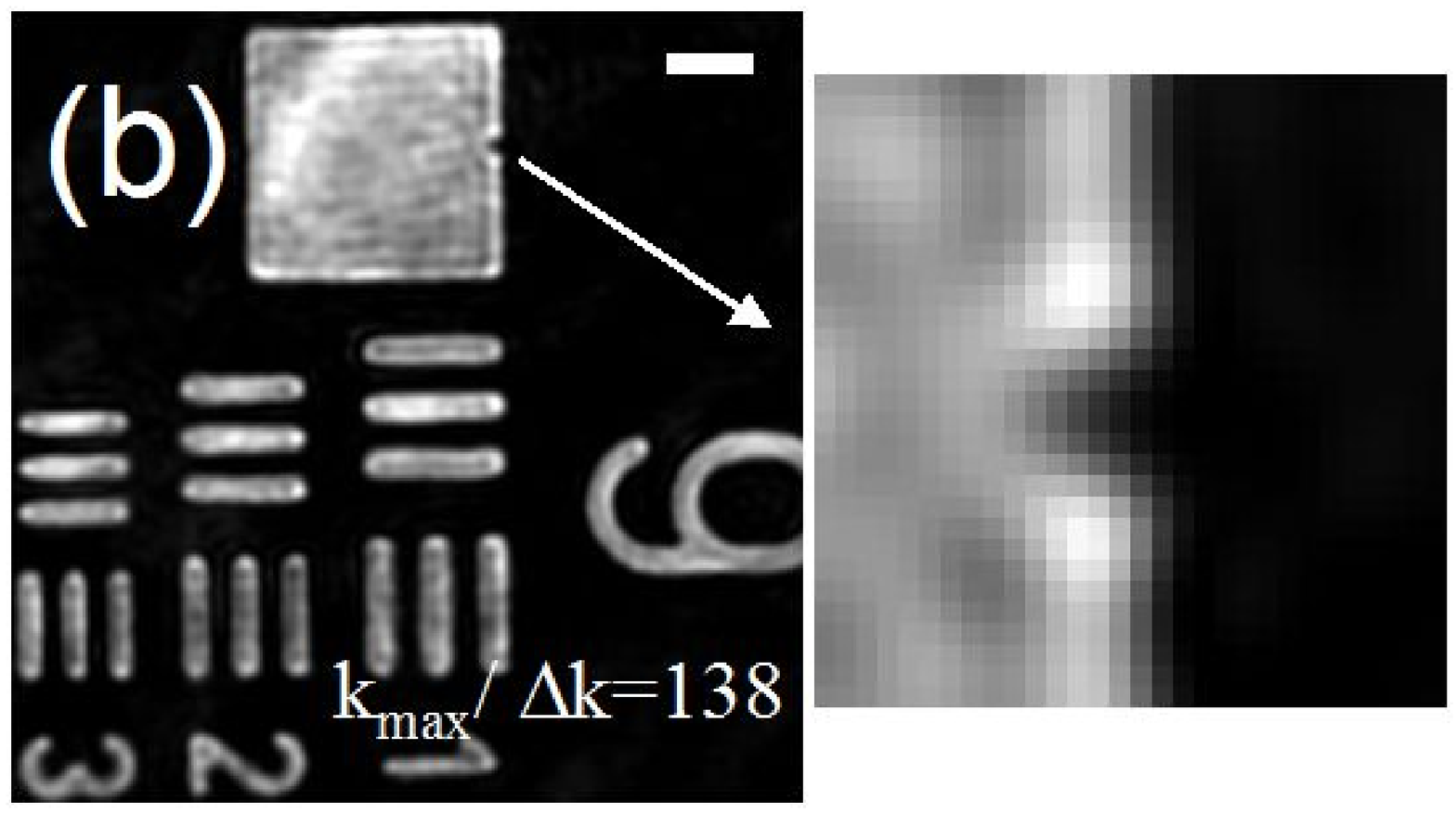}\\
   \includegraphics[width=7cm]{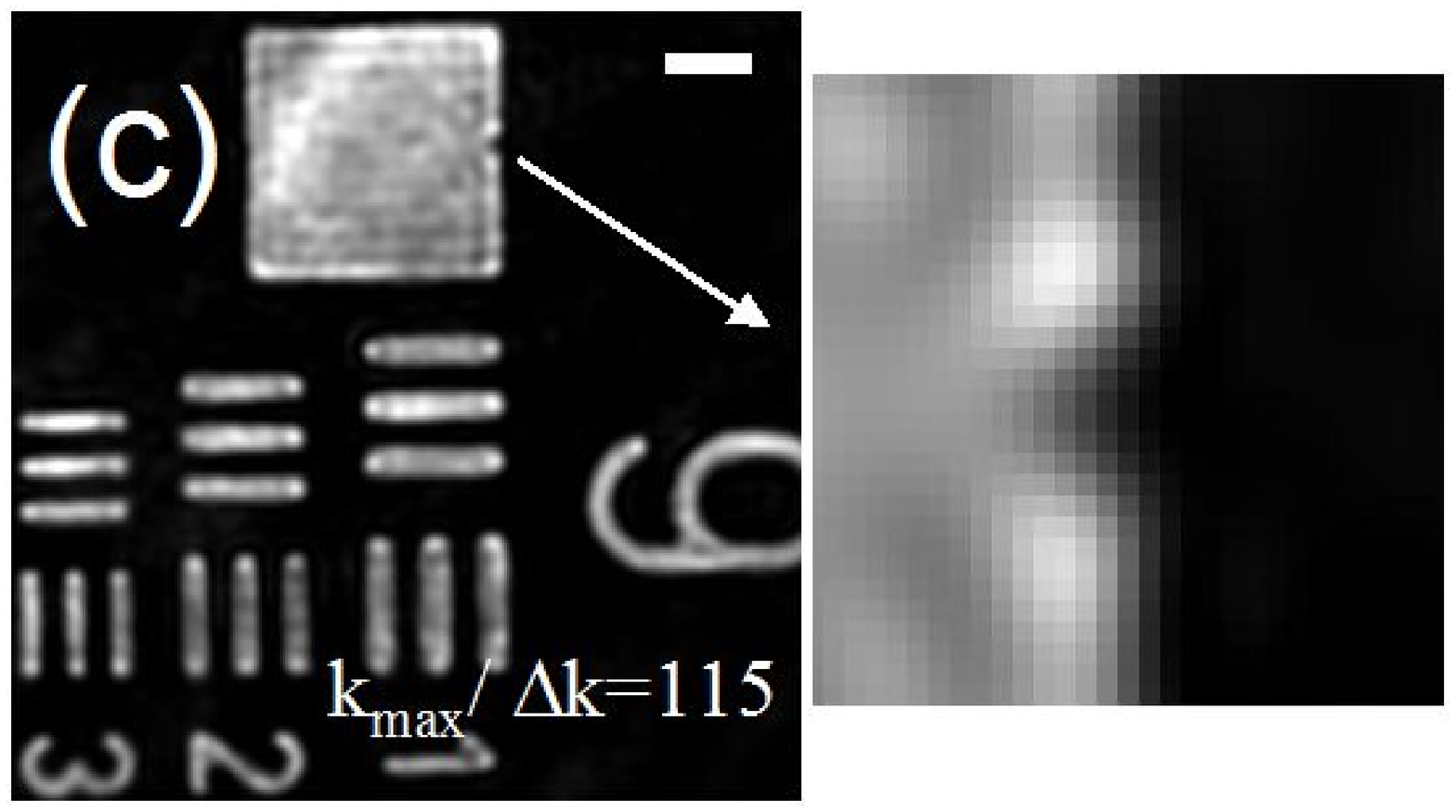}\\
   \includegraphics[width=7cm]{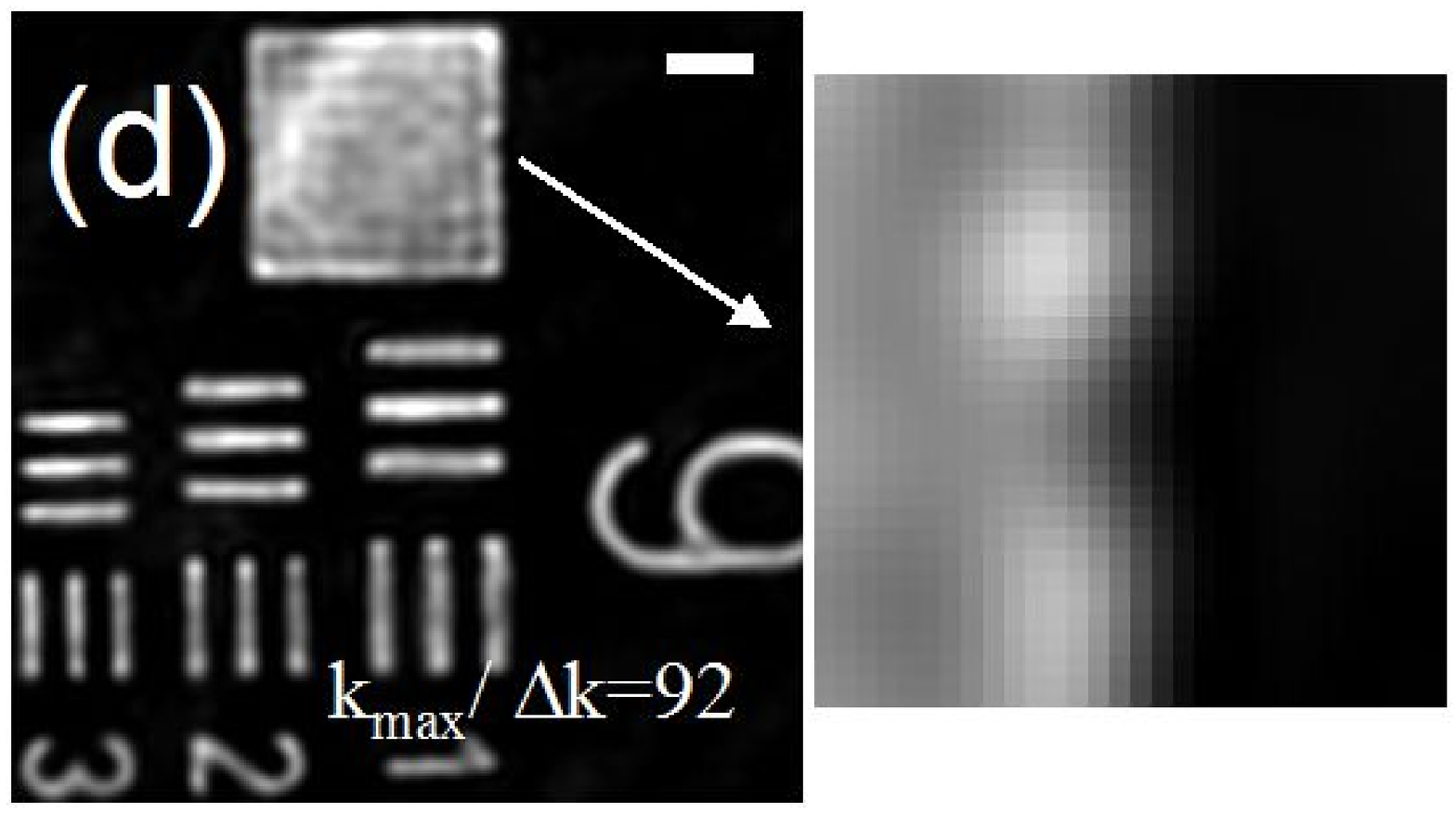}\\
    \caption{ Zooms ($300\times 300$  and $30\times 30$ pixels) of the center of the reconstructed image $|H_{U'}|^2$ ($1024\times 1024$ pixels) of the USAF target obtained for the optimal location $n$=24.  $H_{U'}$ is calculated by  Eq. \ref{Eq_H_U ter}  with $z'_{P'}=-14.34~ \mu$m and $z'_{n}-z'_{P'}= 0.8 ~\mu$m by varying the radius of the selected zone:   $k_{max}/\Delta k=162$    i.e. NA = 1.4 (a), $k_{max}/\Delta k =138$ i.e. NA$\simeq$ 1.2 (b), $k_{max}/\Delta k=115$ i.e. NA$\simeq$ 1.0 (c) and $k_{max}/\Delta k= 92$  i.e. NA$\simeq$ 0.8 (d).  Scale bar is 3 $\mu$m, pixel size is $\Delta x'=89.8$ nm. } \label{Fig_fig13}
  \end{center}
\end{figure}

In order to quantify the resolution we have performed the reconstruction by varying the radius $k_{max}$ of the cropped zone. By the way, we have modified the equivalent  numerical aperture (NA) of microscope objective. To simplify the discussion, we have considered that NA and $k_{max}$ are roughly  proportional.
Figure \ref{Fig_fig13}  shows the images obtained with the USAF target  in position $n=24$ close to the optimal plane P' for   $k_{max}/\Delta k=162$    i.e. NA = 1.4 (a), $k_{max}/\Delta k =138$ i.e. NA$\simeq$ 1.2 (b), $k_{max}/\Delta k=115$ i.e. NA$\simeq$ 1.0 (c) and $k_{max}/\Delta k= 92$  i.e. NA$\simeq$ 0.8 (d).
As expected, the resolution decreases with NA.  Comparison between  Fig.\ref{Fig_fig12} and Fig.\ref{Fig_fig13} quantifies the resolution degradation observed when $ | z'- z'_{P'} | $ increases.
\begin{itemize}
  \item For $| z'- z'_{P'}|\leq  25 ~\mu$m,  the resolution seen on  Fig.\ref{Fig_fig12} (b) et (d) is excellent and remains comparable to the resolution obtained for NA $\simeq$ 1.2 on Fig.\ref{Fig_fig13} (b).
  \item For  $| z'- z'_{P'}| \simeq  50 ~\mu$m   the resolution seen on Fig.\ref{Fig_fig12} (a)  deteriorates slightly. It is lower than that obtained for NA $\simeq$ 1.2  on Fig.\ref{Fig_fig13} (b), but noticeably  better than that the one obtained for NA $\simeq$ 1.0 on Fig.\ref{Fig_fig13} (c).
\item For  $| z'- z'_{P'}| \le 5 ~\mu$m   the resolution seen on Fig.\ref{Fig_fig12} (d)  is the same than for  NA = 1.4 on Fig.\ref{Fig_fig13} (c).

\end{itemize}

Note that the low degradation of the resolution, that is observed on  Fig.\ref{Fig_fig12},  is obtained only if the off-axis tilt parameters   $dk_x, dk_y  $ are properly adjusted. To avoid spherical aberration with high NA objective, it is indeed necessary that the reconstruction $ z $ axis exactly coincides  with microscope objective symmetry axis.
By adjusting  $dk_x, dk_y  $  so that the $ x', y' $  position  of the reconstructed image do not depend on  the $z' $ position of the USAF target, we have aligned the $ z $ reconstruction axis  with the $ z $ axis of translation of the object. As our experimental device is constructed by modifying a commercial microscope, the calibration procedure makes the $z $  axis  coincident with the   MO optical axis, and spherical abberations are minimized.
%

%
%
%
%

\section{Conclusion}

We have proposed an holographic microscopy reconstruction method compatible with  high numerical aperture microscope objective MO. The key idea is to apply in  the  plane of the camera a phase mask (RM+NL) that transforms the optical elements of the  holographic setup into an afocal device. This mask  makes also  the standard phase corrections: phase curvature induced by the microscope objective,   phase curvature  of the reference, and  off axis  phase.

The reconstruction can be then make in 3 steps:
\begin{itemize}
  \item The hologram is first propagated in air from the camera plane C to the image P'' of the MO optimal plane P' (in the object half space)  by the afocal device. Since P'' is at finite distance, the propagation is made by the angular spectrum method. Since propagation is made in the image half space with small  angles   and large pixels in  $x$ and $y$, the calculation can be made with a quadratic kernel.
  \item The hologram in the MO optimal plane P' is then calculated from the hologram in plane P'' by changing the scale of the coordinates $x, y, k_x$ and $k_y$.
  \item To the end, the hologram is propagated  from P' to the object by the angular spectrum method.
  Since propagation is made in the object  half space with large angles   and small pixels, this second propagation must   be calculated  with the exact kernel.
\end{itemize}

This 3 steps reconstruction can be formally written in 2 steps with a propagation of the hologram  from the camera image plane C' to object optimal plane P' with a quadratic kernel, followed by a propagation from P' to the object with the exact kernel.

The proposed reconstruction is parameterized by few calibrations parameters that   are
\begin{itemize}
  \item the focal length $r''$ of  RM+NL  phase mask,
  \item the off axis    translation   $dk_x ,
  dk_y   $ in Fourier space,
  \item the microscope enlargement factor $G$ that is needed to change the scale of the coordinates,
  \item and the location $z'_{P'}$ of the optimal plane P' in the object half space.
 \end{itemize}
These parameters are easily determined  by imaging an object like an USAF target  for different positions along $z'$.
Once the calibration is done, the reconstruction can be made with an object located in any plane $z'$.

The reconstruction method has been validated with a USAF target that is imaged with a  NA=1.4 microscope objective.
%
%
With proper calibration, near-optimal resolution over a wide range of sample locations along the $z$
axis can be obtained. We get a resolution better than the one obtained with NA $\simeq $ 1.0 within a
range of $\pm$ 50 $\mu$m in $z$, and better than the one obtained with NA $\simeq$ 1.2 for range of $\pm$ 25 $\mu$m.
To get the best results with high NA objective, it is of paramount importance to correctly adjust the reconstruction $z$ axis to make it precisely coincides with the  MO optical axis in order to avoid spherical abberation. This can be done easily if the translation of the object made in calibration coincide with the microscope objective optical axis.

We acknowledge ANR Blanc Simi 10 (n 11 BS10 015 02) grant and Labex Numev (convention ANR-10-LABX-20) grant for funding.

\bibliographystyle{unsrt}


\end{document}